\documentclass[fleqn,usenatbib]{mnras}

\usepackage{newtxtext,newtxmath}
   
\usepackage[T1]{fontenc}
\usepackage{ae,aecompl}
\usepackage{graphicx}   
\usepackage{amsmath}    
\usepackage{amssymb}    

\usepackage{makecell}

\title[Search for radio bursts in M82]{A search for millisecond radio bursts from Messier~82}

\author[Paine et al.]{
S.~Paine$^{1,2}$,
T.~Hawkins$^{1}$, 
D.~R.~Lorimer$^{1,2}$,
J.~Stanley$^{1}$,
J.~Kania$^{1,2}$,
F.~Crawford$^{3}$ and
N.~Fairfield$^{4}$
 \vspace{0.5cm} \\
$^{1}$Department of Physics and Astronomy, West Virginia University, Morgantown, WV 26506-6315, USA\\
$^{2}$Center for Gravitational Waves and Cosmology, Chestnut Ridge Building, Morgantown, WV 26505, USA\\
$^{3}$Department of Physics and Astronomy, Franklin \& Marshall College, P.O. Box 3003, Lancaster, PA 17604, USA\\
$^{4}$Amateur Astronomer
}

\date{Accepted 2024 January 30. Received 2024 January 26; in original form 2023 August 12}

\pubyear{2023}

\begin{document}
\label{firstpage}
\pagerange{\pageref{firstpage}--\pageref{lastpage}}
\maketitle

\begin{abstract}
Fast radio bursts (FRBs) are short-duration radio pulses of cosmological origin. Among the most common sources predicted to explain this phenomenon are bright pulses from a class of extremely highly magnetized neutron stars known as magnetars. Motivated by the discovery of an FRB-like pulse from the Galactic magnetar SGR~1935+2154, we searched for similar events in Messier 82 (M82). With  a star formation rate 40 times that of the Milky Way, one might expect that the implied rate of events similar to that seen from SGR~1935+2154 from M82 should be 40 times higher than that of the Milky Way. We observed M82 at 1.4~GHz with the 20-m telescope at the Green Bank Observatory for 34.8 days. While we found many candidate events, none had a signal-to-noise ratio greater than 8. We also show that there are insufficient numbers of repeating low-significance events at similar dispersion measures to constitute a statistically significant detection. From these results, we place an upper bound for the rate of radio pulses from M82 to be 30~year$^{-1}$ above a fluence limit of 8.5~Jy~ms. While this is less than 9 times the  rate of radio bursts from magnetars in the Milky Way inferred from the previous radio detections of SGR~1935+2154,  it is possible that propagation effects from interstellar scattering are currently limiting our ability to detect sources in M82. Further searches of M82 and other nearby galaxies are encouraged to probe this putative FRB population.
\end{abstract}


\section{Introduction}

Fast radio bursts (FRBs) are bright millisecond pulses of radio emission which occur uniformly on the sky at the rate of a few thousand per day \citep{2007Sci...318..777L,2013Sci...341...53T}. Observed FRBs are typically short in duration, with the majority in the range 1--10~ms {\citep{FRBSTATS, CHIMECat}}. Peak flux densities vary, but are typically in the range 0.1--10~Jy {\citep{FRBSTATS, CHIMECat}}. Although sky locations are currently not precise enough to allow identifications of counterparts for most of the known sources, so far 43 FRBs have been confidently associated with host galaxies, clearly establishing them as a cosmological population. Recent reviews can be found in \citet{2022A&ARv..30....2P} and \citet{2022Sci...378.3043B}.

Among the currently observed sample of 789 FRBs\footnote{For an up-to-date list, see https://blinkverse.alkaidos.cn} are 65 repeating sources. Possible explanations as to why the vast majority of FRBs have so far been observed only once are that {some} FRBs repeat on a  timescale {longer than currently probed} and/or there are multiple distinct FRB populations. Recent results from the CHIME/FRB {collaboration suggest} that repeating FRBs and non-repeating FRBs are morphologically distinct, with repeating FRBs typically exhibiting longer pulses than {non-repeating FRBs} \citep{Pleunis2021,Zhong2022}. Repeating FRBs also have smaller emission bandwidths, {typically} 100--200~MHz, while non-repeating FRBs typically occupy the entire CHIME bandwidth  {\citep[400--800~MHz;][]{CHIMECat}.} 

On April 28, 2020, an FRB-like pulse (hereafter FRB~20200428A) was detected from a Galactic magnetar \citep{CHIME2020, Bochenek2020}. The magnetar in question, SGR~1935+2154, was in a period of unusually high X-ray activity at the time \citep{Younes2020,Mereghetti2020}. FRB~20200428A has two components of less than a ms, separated by  29~ms \citep{Bochenek2020, CHIME2020} and the observed dispersion measure (DM) is 332.7~${\rm cm}^{-3}~\text{pc}$. Since the radio phenomenology shares many similarities with FRBs, a natural conclusion to draw from this clear association is that FRB~20200428A is an example of a Galactic FRB, albeit with a lower luminosity due to its closer proximity to Earth. As discussed by \citet{Bochenek2020}, if this hypothesis is correct, and magnetar flares can produce at least some FRBs, then excellent targets for further searches are nearby galaxies with high levels of star formation. One such example is the starburst galaxy Messier 82 (hereafter M82), in which the star formation rate is approximately 40 times that of the Milky Way \citep{2020PASP..132c4202B,2021ApJ...908...61K}.

Subsequent detections of fainter radio pulses from SGR~1935+2154 were made by \citet{2021NatAs...5..414K} using 20--30~m class telescopes at Westerbork, Torun and Onsala, and by {\citet{2020ATel13699....1Z}} using the Five Hundred Metre Aperture Spherical Telescope (FAST), \citet{2020ATel14074....1G}, \cite{2022ATel15681....1D}, and
\citet{2022ATel15792....1P} using CHIME, \citet{2020ATel14186....1A} at 111~MHz using the BSA, \citet{2022ATel15697....1M} using the Green Bank Telescope and by \citet{2022ATel15707....1H} using the Yunnan 40~m telescope in China. Marginal detections of pulses were also seen using the Northern Cross Radio Telescope \citep{2020ATel13783....1B}.

\citet{2023arXiv230716124Z} announced the detection of radio pulsar-like emission with FAST from  SGR~1935+2154 some five months after the initial outburst of FRB~20200428A. Their observations of almost 800 pulses over a period of two weeks make a strong case for two states in this magnetar: one in which bright outbursts occur at random pulse phases (possibly during periods of intense magnetospheric activity), and another in which regular pulses are emitted within a narrow range of pulse phases. This would strengthen the magnetar--FRB connection and  account for the lack of periodicity so far found in any repeating FRBs \citep[see, e.g.,][]{2022RAA....22l4004N}.

Motivated by the prospect of probing FRBs and magnetars in a new way, we have carried out a dedicated survey of M82 which has good sensitivity to radio pulses from sources like SGR~1935+2154. The rest of this paper is structured as follows. In \S 2, we discuss the predictions from the magnetar hypothesis in more detail. In \S 3, we describe our observations using the 20-m telescope at Green Bank. In \S 4, we summarize our results and discuss the main findings. Finally, in \S 5, we present our conclusions.

\section{The FRB--magnetar connection}

Though the origins of FRBs are still unknown, most theories agree that FRBs arise from compact objects which have access to significant resources of energy necessary to produce the observed emission \citep{Platts_2019}. Magnetars, neutron stars with magnetic fields  around $10^{15}$~G \citep[for a comprehensive review, see][]{Kaspi2017}, fit that description, and are strong candidates as sources for at least some fraction of the FRB phenomenon. Magnetars distinguish themselves from their radio pulsar counterparts in that their dominant energy resource is in their magnetic fields. In dramatic fashion, magnetars release some of their energy in the form of bursts that are visible by instruments across the electromagnetic spectrum, including in gamma rays and X-rays. The FRB-like pulse from SGR~1935+2154 was temporally coincident with a burst of both soft gamma rays and hard X-rays \citep{CHIME2020,Younes2020,Mereghetti2020}. This leads to the possibility that a number of FRBs have higher frequency components that are not detected due to being at extragalactic distances.

\begin{figure} \includegraphics[width=0.51\textwidth]{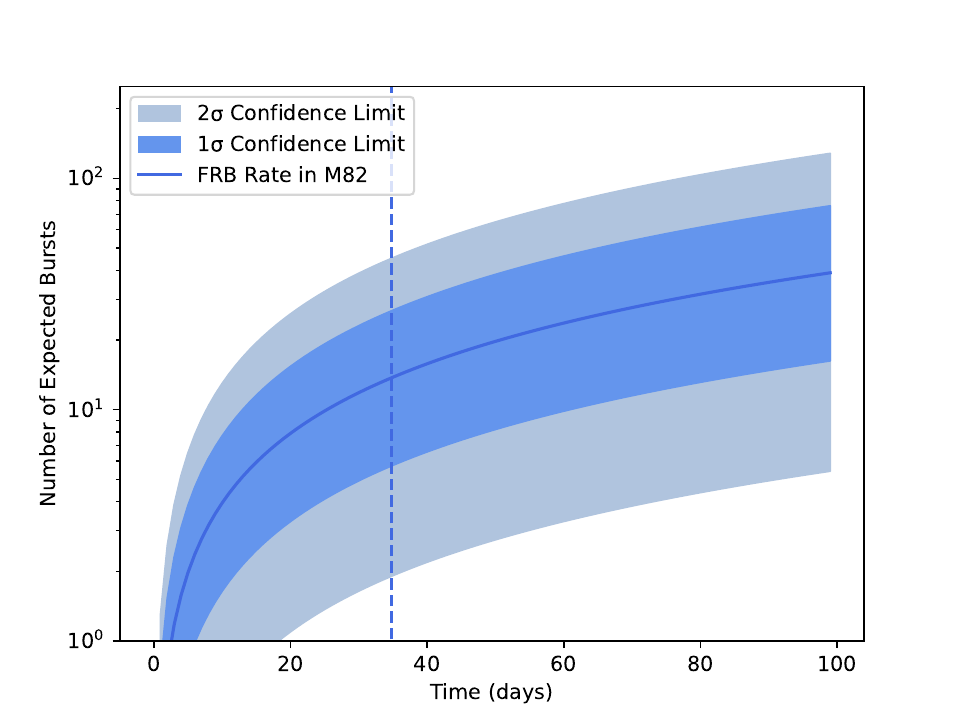}
\caption{The expected number of radio bursts found from M82 as a function of time. The central line represents the STARE2 rate multiplied by the ratio of star formation rates of M82 and the Milky Way. The shaded regions represent $1\sigma$ and $2\sigma$confidence levels.}
\label{fig:numberofbursts}
\end{figure}

SGR~1935+2154 was particularly active during the period in which an FRB-like pulse was detected \citep{CHIME2020}. Assuming that the rate of radio bursts from magnetars in a particular galaxy will scale roughly with star formation rate, M82, a nearby starburst galaxy, should have a higher number of magnetars than other galaxy {types. According} to \citet{Bochenek2020}, the rate of potentially observable FRB-like pulses with fluences above 1.5~MJy~ms $R=3.6^{+3.4}_{- 2.0}$~year$^{-1}$, where the ranges denote a 1$\sigma$ (68\% confidence interval) uncertainty. This result follows from a Poissonian analysis \citep[for details, see, e.g.,][]{2016MNRAS.455.2207R}, where the probability density 
\begin{equation}
    P(R) = RT \exp (-RT),
\end{equation}
with $T=0.468$~yr representing the total time observed by STARE2. Taking the 68 and 95\% confidence intervals, we then multiply by a factor of 40 to estimate the FRB-like burst rate in M82. Assuming such events are above the detection threshold of a given observing system (see below), the number of FRBs we expect to find after a length of time within those bounds is plotted in Fig.~\ref{fig:numberofbursts}. At 34.8 days, we expect between 6 and 26 events. The shaded regions represent the possible range of bursts found by that time. We anticipate a detection of an FRB-like burst  within the first $\sim$20 days of observation.

To estimate the detectability of FRB-like bursts with the 20-m telescope, we take the fluence of FRB~20200428A found by \citet{Bochenek2020} to be 1.5~MJy~ms and a measured width of 0.6~ms. At the distance of M82 of 3.66~Mpc \citep{2013AJ....146...86T} compared to the 9~kpc distance to SGR~1935+2154 \citep{2020ApJ...898L...5Z}, we estimate a fluence of 9.4~Jy~ms if this event came from M82. Using the radiometer equation \citep{Lorimer2004} and adopting the parameters of the 20-m telescope given in the next section, we estimate a signal-to-noise ratio, S/N~$\sim 11$. While this is sufficient to begin to test the hypothesis presented here, as discussed below, enhanced sensitivity in future would provide more stringent constraints.

\section{Observations and data analysis}

\begin{figure*} \includegraphics[width=0.99\textwidth]{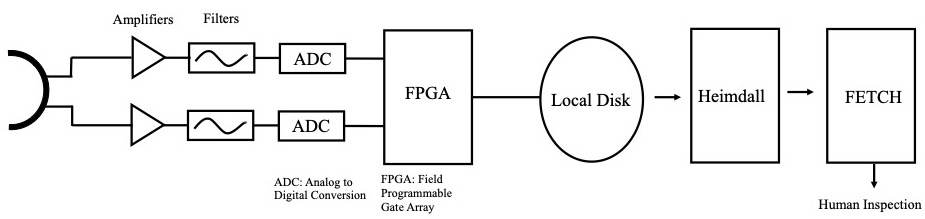}
\caption{Schematic of the data acquisition and analysis pipeline, from the receiver to the final data outputs.}
\label{fig:pipeline}
\end{figure*}

We carried out observations of M82 using the 20-m telescope at the Green Bank Observatory. The telescope is available for use via the remote telescope operating system Skynet \citep{2013AAS...22132807L}. The telescope receiver operates at a central frequency of 1.4~GHz using a cryogenically cooled system which collects data from two orthogonal polarization channels. A summary of the main characteristics in the system is given in Table~\ref{tab:importvalues}. Between October 2020 to January 2022, we scheduled observations of a maximum of {twelve} hours per epoch. In total, we obtained 34.8 days of on-source time on M82. 

After amplification and filtering, each of the dual polarization channels is digitized and combined in a dedicated field programmable gate array which samples the observing band every 131~$\mu$s. The resulting observation files are saved to disk in {\tt psrfits} format \citep{2010PASA...27..104V}. By default, these files contain full Stokes parameters and additional (unused) frequency channels. Within the nominal 125~MHz band, which is split into 256 channels, due to the presence of strong interference, a filter was applied which limits the usable part of the spectrum to the 80 MHz between $\sim$1360-1440 MHz. In this experiment, as the telescope is not fully calibrated, and we are only interested in Stokes I, the {\tt psrfits} files are combined into a single {\tt filterbank} file \citep{2011ascl.soft07016L} with 256 individual frequency channels. Because of the presence of the filter, the unused frequency channels are set to zero so as not to introduce any noise into the downstream analysis. Those files are then sent to HEIMDALL and FETCH as discussed below. The pipeline is shown schematically in Fig.~\ref{fig:pipeline}. 

\begin{table}
    \centering
    \begin{tabular}{ c c c } 
 Parameter & Value & Unit \\ 
 \hline
 Telescope gain, $G$ & 0.086 & K $\text{Jy}^{-1}$ \\
 Total bandwidth, $\Delta \nu$ & 125 & MHz \\
 Usable bandwidth, $\Delta \nu$ & 80 & MHz \\
 Number of channels, $n$ & 256 &  \\
 Channel bandwidth, $\Delta \nu_{\text{chan}}$ & 0.488 & MHz \\
 System temperature, $T_{\text{sys}}$ & 40 & K \\
 Center frequency, $\nu_{0}$ & 1.4 & MHz \\
 Sampling interval, $t_{\text{samp}}$ & 131.07 & $\mu$s \\
 
\end{tabular}
    \caption{System values for the 20-m telescope at Green Bank Observatory.}
    \label{tab:importvalues}
\end{table}

\begin{figure}
    \centering
    \includegraphics[width=0.5\textwidth]{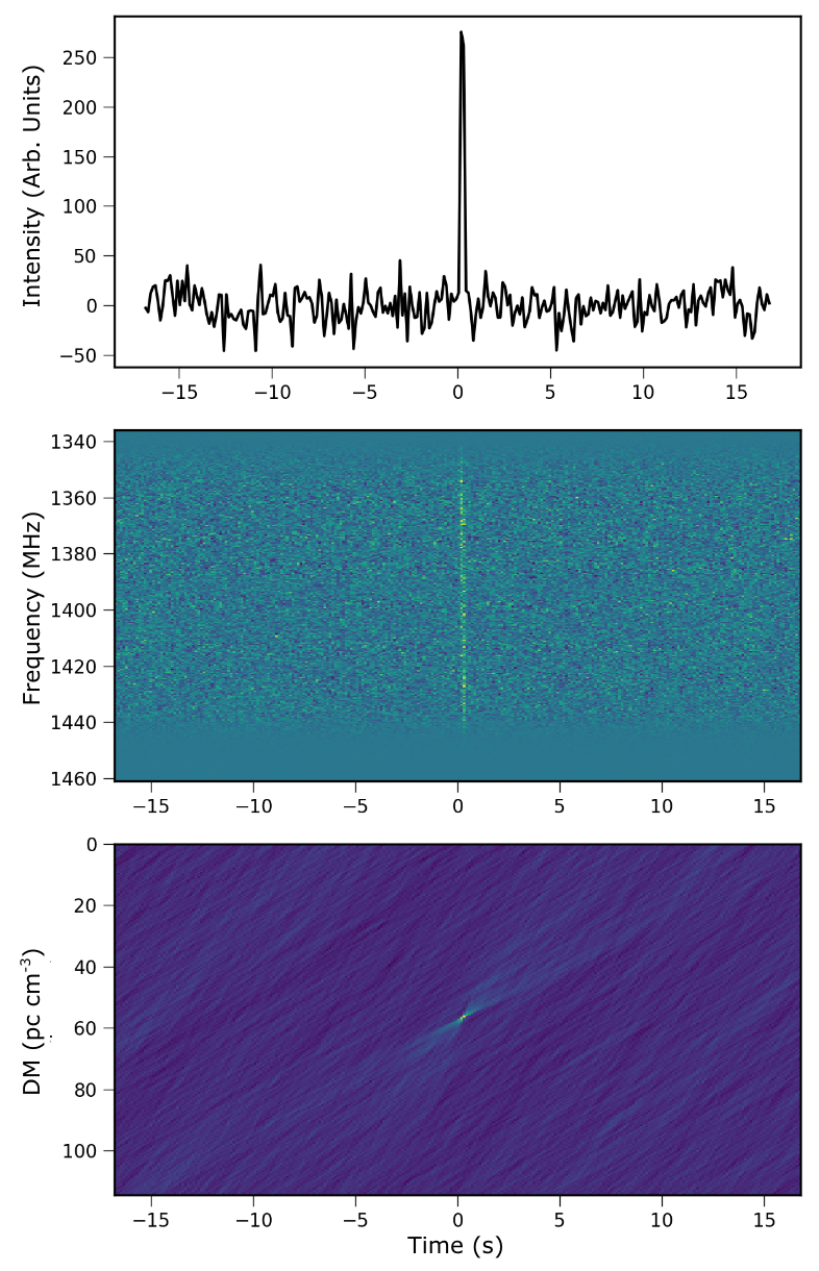}
    \caption{A sample pulse from the Crab Pulsar. The top panel shows dedispersed intensity versus time, the middle panel shows intensity versus dedispersed frequency and time, and the bottom panel shows DM versus time. This particular event has a DM of 57.17 cm$^{-3}$~pc and a S/N of 25.}\label{fig:crabpulse}
\end{figure}

The HEIMDALL\footnote{https://sourceforge.net/projects/heimdall-astro} software package \citep{Barsdell2012} performs incoherent dedispersion and single-pulse searches on graphical processing units.  HEIMDALL generates dedispersed time series from the data using 471 trial DMs in the range 0--10,000~cm$^{-3}$~pc. For each DM trial, individual pulses are sought via a matched filtering process in which box-car kernels are used which are optimized to find pulses of varying widths in the range 0.131--67.1 ms. HEIMDALL also carries out some radio frequency interference removal using zero-DM filtering  \citep{2009MNRAS.395..410E}. If an excess of power is found with a S/N of at least 6, HEIMDALL flags the event as a candidate pulse. An example pulse from a test observation of the Crab pulsar is shown in Fig.~\ref{fig:crabpulse}. This is a standard image format that
is generated for all candidates from HEIMDALL and shows the dedispersed pulse (top), dedispersed frequency versus time (middle), and the search for the pulse in DM space (bottom).

These data were searched for dispersed pulses using HEIMDALL \citep{Barsdell2012}. To optimize the process in which FRB candidates from HEIMDALL are assessed, we used FETCH \citep[Fast Extragalactic Transient Candidate Hunter][]{Agarwal2019}, an open-source machine learning platform originally developed for use on the GREENBURST experiment \citep{2019PASA...36...32S,2020MNRAS.497..352A}. FETCH uses convolutional neural networks to analyse images of the form shown in Fig.~\ref{fig:crabpulse}.  To select the most likely pulses of astrophysical origin, FETCH has been extensively trained on data acquired from surveys and observations with multiple telescopes. We used FETCH in its
default training set, model A. While it is known \citep[see, e.g.,][]{2023MNRAS.520.2281N} that this model can miss weak events (S/N~$<10$), it is adequate for the purposes of this search where we are searching for events of greater statistical significance. The low-significance events from M82 reported in this work are much closer to the noise level than the Crab pulsar test observation shown in Fig.~\ref{fig:crabpulse}. 

\begin{figure}
    \centering
    \includegraphics[width=0.5\textwidth]{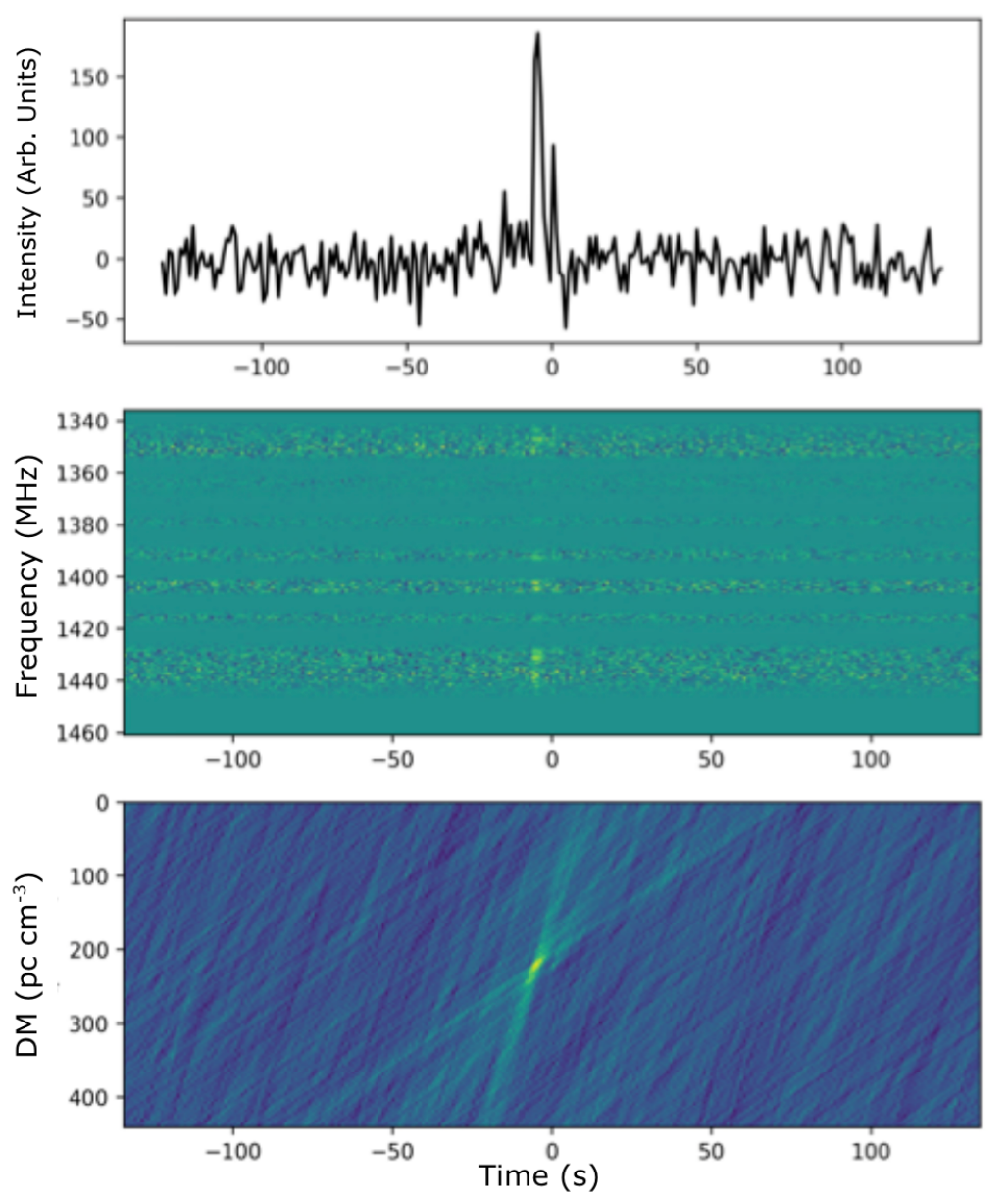}
    \caption{A sample pulse from FRB 20220912A which was detected by the same search pipeline used for the 20-m observations of M82.}\label{fig:frbpulse}
\end{figure}

To further validate our data collection system, we observed FRB~20220912A, a recently discovered repeating FRB \citep{FRBDisc} from the CHIME/FRB experiment which has subsequently been observed by a number of other telescopes, including FAST \citep{Zhang2023}. We observed FRB~20220912A for 10 hours as part of a larger multi-telescope campaign reported elsewhere \citep{2023ATel16196....1D}. For the purposes of this paper, we note that FRB~20220912A is the first confirmed FRB detection with the 20-m telescope at the Green Bank Observatory. As it is similar in brightness to putative FRBs in M82, FRB~20220912A  serves as an excellent test source. We detected three clear pulses from FRB~20220912A \citep{2023ATel16196....1D}. One of them is shown in Fig.~\ref{fig:frbpulse}. The pulse has a S/N 28.8 and a DM of 220.5~cm$^{-3}$~pc. This DM is consistent with that found by \citet{FRBDisc} and \citet{Zhang2023}.

\section{Results and Discussion}

From a total of 835.3 hours of observation time on M82, summarized in Table 2, 291 single-pulse candidates were detected. Most of the reported S/Ns were around 6, with the highest being 7.4. These are all below our formal threshold of 8.5~Jy~ms which would be expected for events with S/Ns of 10 or greater. In the subsections below, we discuss the implications of these results.

\subsection{Low significance pulses}

\begin{figure}
    \centering
    \includegraphics[width=0.52\textwidth]{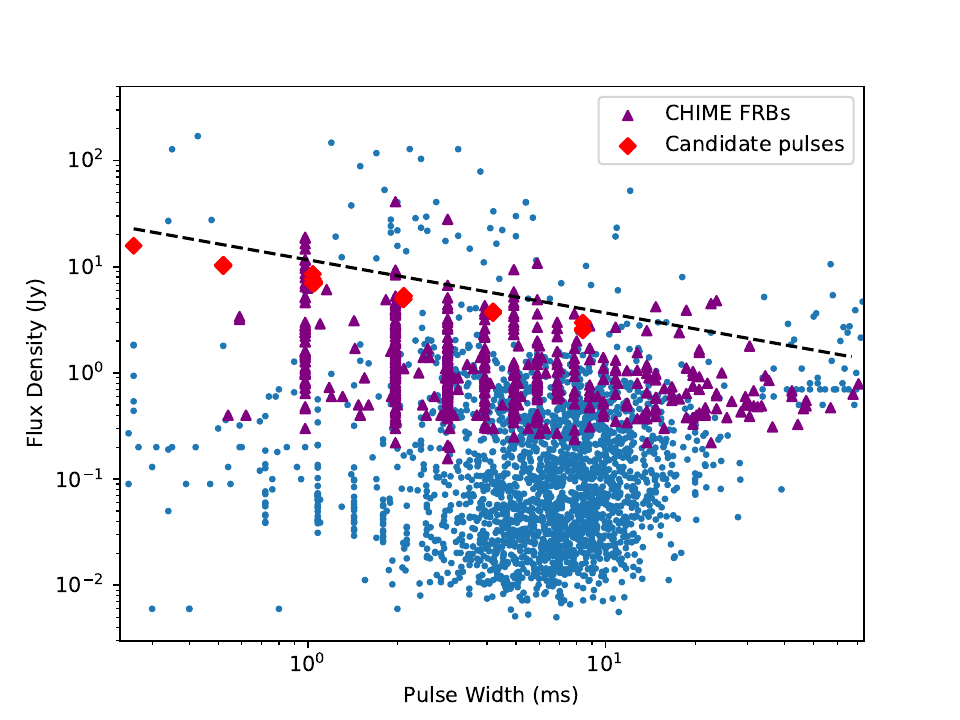}
    \caption{Flux density versus pulse width for known FRBs and our candidates. Known FRBs are in blue \citep{2023Univ....9..330X}, the black dashed line shows the limit corresponding to S/N of 10 which extends over the range of pulse widths searched. The quantization seen for some pulse widths is due to the limited measurement precision of those entries in the database.}\label{fig:pwvfd}
\end{figure}

From our test observations, and prior experience in other single-pulse search experiments \citep[see, e.g.,][]{2022MNRAS.509.1929P}, it is difficult to establish the validity of a single pulse when its S/N is below {8}. Such candidates are generally weak and hard to discern features in the diagnostic plots. As recently demonstrated by \citet{2022MNRAS.509.1929P}, a comparable number of low S/N pulses can be found by de-dispersing the data with negative {DMs.}

To demonstrate the sensitivity of our experiment, we have converted the S/Ns of our most promising candidates into peak flux densities using the radiometer equation \citep[for further discussion, see][]{Golpayegani_2019} and plotted them as a function of pulse width in Fig.~\ref{fig:pwvfd}. For context, we also show the currently known FRBs \citep{2023Univ....9..330X}. These lie below the approximate detection threshold shown for S/N of 10. Our experimental setup is only sensitive to the sample of brighter FRBs with flux densities greater than approximately 10.4~Jy$/\sqrt{W}$ for pulse widths, $W$, in ms. As shown in Fig.~\ref{fig:pwvfd}, this sample makes up a smaller but non-negligible fraction of the population. Our intention with this comparison is to show that, in the absence of any significant propagation effects discussed below, the 20-m telescope is able to detect the brighter end of the FRB sample.

\subsection{Repeating pulses}

One way to establish the astrophysical reality of a candidate is the detection of a repeating source which emits multiple pulses at the same DM. In such cases, the probability of non-astrophysical events occurring by chance will decrease when more pulses have the same DM. In our list of candidates, we found two DMs each have four low S/N repetitions. These DMs are 719 cm$^{-3}$~pc and 216 cm$^{-3}$~pc. While these DMs are likely on the low end of any sources in M82 (see \S \ref{sec:limits}), we now quantify the significance of these results. 

Using a Poissonian framework, the probability of finding $n$ pulses at the same DM from a sample of $N$ total pulses found in a search over $T$ DM trials,
\begin{equation}
\label{eq:poisson}
    P(n|N,T) = \frac{(N/T)^n \exp (-N/T)}{n!}.
\end{equation}
Summing Eq.~\ref{eq:poisson} gives the probability of obtaining greater than or equal to $n$ events by chance,
\begin{equation}
    p = \sum_{i=n}^{\infty} \frac{(N/T)^i \exp (-N/T)}{i!}.
\end{equation}
The complement of this result { is} the confidence level { in a detection} of $n$ or more { real} pulses at the same DM, 
\begin{equation}
   C = 1-p = \sum_{i=0}^{i=n-1} \frac{(N/T)^i \exp (-N/T)}{i!},
\end{equation}
where as $n$ increases,  $C$ also increases. In our case, with $n=4$ from a sample of $N=291$ pulses with $T=462$ DM trials, we find $C=99.6$\%, i.e., an equivalent gaussian significance of only 3$\sigma$. To reach a 5$\sigma$ significance, we would need at least eight pulses at the same DM. We therefore conclude that the four pulses detected in each of two DM trials are not statistically significant with expectations for an astrophysical source and are instead entirely consistent with a random sampling from our candidates.

\subsection{Constraining the rate of FRB-like pulses from M82} \label{sec:constrain}

Since none of the candidate events discussed above can be confidently classified as astrophysical in origin, this null result is in tension with our hypothesis: after 34.8 days, as shown in Fig.~\ref{fig:numberofbursts}, we would have expected about 10 events above our detection threshold from M82. Without a single detection, we must reexamine our assumptions to draw new conclusions.

In the Poissonian regime, given a burst rate $R$, the probability of finding no events after a time $T$ can be written \citep[see, e.g.,][]{2016MNRAS.455.2207R} simply as 
$P(0|R,T) = \exp (-RT)$. Setting this as the likelihood in a Bayesian analysis with a flat prior in $R$ gives the same decaying exponential functional form for the posterior PDF of $R$. The formal upper limit on $R$ for some confidence level $C$ is then
\begin{equation}
    R = - \frac{\ln(1-C)}{T}.
\end{equation}
Setting $T=34.8$~d and $C=0.95$, the 95\% upper bound on the rate of FRB-like pulses, given our lack of detections with the 20~m is 0.09~day$^{-1}$ or about 30~year$^{-1}$ with fluences above 8.5~Jy~ms. We can use this result to place an upper bound on the relative star formation rate for M82, based on \citet{Bochenek2020} where the rate of FRB-like pulses was found to be $R_{\rm STARE2}$. We can express this condition as
\begin{equation}
\label{mwrate}
   \frac{\text{SFR}_{\text{M82}}}{\text{SFR}_{\text{MW}}} = \frac{{R}_{\text{M82}}}{{R}_{\rm STARE2}},
\end{equation}
where SFR$_{\text{M82}}$ is the star formation rate of M82, SFR$_{\text{MW}}$ is the star formation rate of the Milky Way, and $R_{\text{M82}}$ is the rate in M82 that we have constrained to be less than 30 per year. Taking the nominal value of $R_{\rm STARE2}$ to be 3.6~year$^{-1}$, we find that ${\text{SFR}_{\text{M82}}}/{\text{SFR}_{\text{MW}}}$ < 9. This rate blatantly contradicts \citet{Barker2008}, who find ${\text{SFR}_{\text{M82}}}/{\text{SFR}_{\text{MW}}}$ = 40.

\subsection{Observational limitations of the current study} \label{sec:limits}

In view of the apparent tension between our formal upper limit on SFR versus that found by \citet{Barker2008}, is important to investigate the deleterious impacts of propagation effects on our sensitivity which might explain our lack of detections with the 20~m. As shown by \citet{1989ApJ...345..163P} from hydrogen recombination line measurements, the average electron density in M82 is likely to be $\sim 30$~cm$^{-3}$. A typical DM contribution from a line of sight piercing a few kpc into M82, then, would be around several thousand cm$^{-3}$~pc. Although we searched up to DMs of $10^4$~cm$^{-3}$~pc, propagation effects from scattering and dispersion could significantly hinder the detection of any pulses. 

To quantify this, consider three example DMs of 1000, 3000 and 9000~cm$^{-3}$~pc. Using the scaling law between DM and scattering time derived from Galactic pulsars given in Eq.~8 of \citet{2022ApJ...931...88C}, we find scattering times of 56~ms, 7.8~s and 1000~s respectively. Correcting for the geometrical effect of having a scattering screen close to the source \citep[see, e.g., Eq.~1 of][]{2013MNRAS.436L...5L} we find 0.7~ms, 93~ms and 13~s, respectively. Similarly, the dispersion broadening across individual frequency channels are, respectively, 1.5, 4.4~ms and 13~ms, for DMs of 1000, 3000 and 9000~cm$^{-3}$~pc. While pulses with DMs at the low end of this range would be detectable, the broadening particularly from scattering is unavoidable at the higher DMs. From these calculations we conclude that our current experiment is not fully sensitive to the range of DMs anticipated in M82.

\subsection{Constraints on radio activity from GRB~231115A}

After the initial submission of this paper, multiple high energy observatories reported the detection of a gamma-ray burst, GRB~231115A, from the direction of M82 \citep{2023GCN.35035....1F,2023GCN.35037....1M,2023GCN.35038....1B}. The short duration (30~ms) of this event and its energetics favour a magnetar flare origin \citep{2023GCN.35044....1D,2023GCN.35065....1R}. GRB~231115A appears to represent exactly the kind of source that we have been searching for in this work. The positional offset of this from the center of M82, well away from the peak of the $H\alpha$ emission, suggests that any radio emission from this source is unlikely to be significantly affected by interstellar scattering. As described by \citet{2023ATel16341....1C}, GRB~231115A was within the CHIME/FRB field of view during the time of the high-energy flare, setting an upper limit of 260~Jy on the peak flux density in the 400--800~MHz band. Optimal sensitivity of 0.5~Jy was obtained some 80 minutes prior to the event when the source transited through the meridian. CHIME/FRB archival observations also provide upper limits on the source activity over the last two years at the 0.5~Jy level. Our observations complement these and provide upper limits in the 1.4~GHz band at the level of 0.85~Jy for a 10~ms pulse at each of the epochs in Table 2. At the distance of M82, this corresponds to a radio luminosity upper limit of $1.2 \times 10^{28}$~erg~s$^{-1}$~Hz$^{-1}$.

\subsection{Comparison with other work}
\cite{NorthernCross} published a study searching nearby high star formation galaxies for FRBs using the Northern Cross Radio Telescope at 400~MHz, reporting a single pulse coming from the direction of (though likely not associated with) M101. All other nearby galaxies in their sample, including M82, had no statistically significant detections. The study had 184 hours, or 7.67 days, of M82 observations. Using the same framework as in {\S \ref{sec:constrain}}, this translates with 95\% confidence to an upper limit of about 0.4 day$^{-1}$, for events above their nominal fluence limit of 38~Jy~ms at 408~MHz. {Translating} between this lower frequency and our 20-m telescope observations at 1400~MHz assuming a nominal power-law spectral index of --1.4 \citep[typical for pulsars,][]{2013MNRAS.431.1352B} gives an equivalent fluence limit at 1400~MHz of $\sim 7$~Jy~ms, similar to our threshold. 

\section{Conclusions}

We have attempted to detect radio pulses from the starburst galaxy M82 using 35 days of observations at 1.4~GHz with the 20-m telescope at the Green Bank Observatory. Motivated by the discovery of an FRB-like pulse from Galactic magnetar SGR~1935+2154 \citep{Bochenek2020,CHIME2020}, we hypothesised that if the rate of radio pulses from magnetars scales with the star formation rate, around 10 pulses with S/N~$>$~10 would have been seen from M82 during our observations. We detected no such events with S/Ns greater than 8 and, for a S/N threshold of 10, place an upper limit of 30~year$^{-1}$ on pulses with fluences greater than 8.5~Jy~ms.

While our null result indicates a rate that is much lower than expected based on extrapolations from the rate of FRB-like pulses from Galactic magnetars, we consider this is most likely a result of the sensitivity of our experiment being: (i) very close to the fluences expected from M82; (ii) further hindered by scatter broadening as a result of the highly ionized environment in M82. Searches of M82 at significantly higher frequencies would be a way to make progress in this area. Even at 8~GHz, the expected scatter broadening from a source in M82 would be of order 20~ms. \citet{2021PhDT.........3B} predicts that a 20~GHz survey of M82 would require up to a year of observations to detect pulses. Finally, we note that our expectations of the rate of pulses from magnetars in M82 is based on a Poissonian extrapolation of one event from extended observations of SGR~1935+2154. If the underlying distribution of events from sources in M82 is non-Poissonian in nature, then our observing campaign might have missed an outburst, such as GRB~231115A described above. Further monitoring of {M82 could} ultimately lead to more definitive conclusions than possible here.

\section*{ACKNOWLEDGEMENTS}

We thank Will Armentrout and Dane Sizemore for assistance with the 20-m telescope observations and the anonymous referee for very useful comments. Access fees for the telescope were supported by a grant from the Research Corporation for Scientific Advancement. SP and JS acknowledge support from the National Science Foundation (award \#1950617) for the Research Experiences for Undergraduates program at West Virginia University.

\section*{Data availability}

All data available from the authors upon request.




\bibliographystyle{mnras}
\bibliography{refs}




\begin{table}
    \begin{center}
    \begin{tabular}{lll}
    \multicolumn{1}{c}{Date} & \multicolumn{1}{c}{MJD} & \multicolumn{1}{c}{Duration (hr)} \\
    \hline
     2020/10/13 & 59135.8 & 2.0 \\
     2020/10/14 & 59136.3 & 10.0 \\
     2020/10/16 & 59138.2 & 8.0 \\
     2020/10/18 & 59240.4 & 6.0 \\
     2020/10/19 & 59140.3 & 6.0 \\
     2020/10/20 & 59142.6 & 6.0 \\
     2020/10/22 & 59144.1 & 10.0 \\
     2020/10/24 & 59146.0 & 5.6 \\
     2020/10/14 & 59146.2 & 12.0 \\
     2020/10/15 & 59147.0 & 12.0 \\
     2020/10/26 & 59148.2 & 12.0 \\
     2020/10/27 & 59149.5 & 6.0 \\
     2020/10/28 & 59150.6 & 6.0 \\
     2020/11/02 & 59155.6 & 10.0 \\
     2020/11/03 & 59156.2 & 10.0 \\
     2020/11/04 & 59157.1 & 10.0 \\
     2020/11/05 & 59158.2 & 10.0 \\
     2020/11/12 & 59165.1 & 10.0 \\
     2020/11/12 & 59165.5 & 10.0 \\
     2020/11/13 & 59166.1 & 10.0 \\
     2020/11/13 & 59166.8 & 2.0 \\
     2020/11/15 & 59168.7 & 7.9 \\
     2020/11/16 & 59169.6 & 4.0 \\
     2020/11/17 & 59170.1 & 10.0 \\
     2020/11/20 & 59173.8 & 5.4 \\
     2020/11/21 & 59174.0 & 0.6 \\
     2020/11/22 & 59175.6 & 8.7 \\
     2020/11/23 & 59176.0 & 8.6 \\
     2020/11/23 & 59176.6 & 5.4 \\
     2020/11/24 & 59177.2 & 0.2 \\
     2020/11/25 & 59178.1 & 5.4 \\
     2020/11/27 & 59180.7 & 6.4 \\
     2020/11/29 & 59182.1 & 4.9 \\
     2020/12/01 & 59184.2 & 10.0 \\
     2020/12/02 & 59185.6 & 8.5 \\
     2020/12/07 & 59190.9 & 3.3 \\
     2020/12/08 & 59191.6 & 1.4 \\
     2020/12/09 & 59192.5 & 10.0 \\
     2020/12/10 & 59193.2 & 10.0 \\
     2020/12/18 & 59201.9 & 2.8 \\
     2020/12/19 & 59202.0 & 10.0 \\
     2020/12/20 & 59203.6 & 7.5 \\
     2020/12/23 & 59206.2 & 0.6 \\
     2020/12/23 & 59206.6 & 9.0 \\
     2020/12/24 & 59207.8 & 2.7 \\
     2020/12/26 & 59209.5 & 10.0 \\
     2021/01/19 & 59233.1 & 10.0 \\
     2021/01/19 & 59233.9 & 2.2 \\
     2021/01/20 & 59234.8 & 4.5 \\
     2021/01/21 & 59235.0 & 10.0 \\
     2021/01/22 & 59236.6 & 4.8 \\
     2021/01/22 & 59236.8 & 1.2 \\
     2021/01/22 & 59238.9 & 3.3 \\
     2021/01/23 & 59237.0 & 0.8 \\
     2021/01/25 & 59239.8 & 4.0 \\
     2021/01/16 & 59240.7 & 4.0 \\
     2021/01/28 & 59242.1 & 4.5 \\
     2021/03/31 & 59304.8 & 8.0 \\
     2021/01/20 & 59234.8 & 4.5 \\
     2021/01/21 & 59235.0 & 10.0 \\
     2021/01/22 & 59236.6 & 4.8 \\
     2021/01/22 & 59236.8 & 1.2 \\
     2021/01/22 & 59236.9 & 3.3 \\
     2021/01/23 & 59237.0 & 0.8 \\

   \hline
    \end{tabular}
    \end{center}
    \caption{Observing log summarizing all the dates and lengths of observations of M82 using the 20-m telescope.}
    \label{tab:observs}
\end{table}
\begin{table}
    \begin{center}
    \begin{tabular}{lll}
    \multicolumn{1}{c}{Date} & \multicolumn{1}{c}{MJD} & \multicolumn{1}{c}{Duration (hr)} \\
    \hline
     2021/01/25 & 59239.8 & 4.0 \\
     2021/01/26 & 59240.7 & 4.0 \\
     2021/01/28 & 59242.1 & 4.5 \\
     2021/02/05 & 59250.8 & 0.7 \\
     2021/02/08 & 59253.9 & 6.0 \\
     2021/03/31 & 59304.8 & 8.0 \\     2021/04/03 & 59307.8 & 10.0 \\
     2021/04/26 & 59330.5 & 0.0 \\
     2021/04/26 & 59330.5 & 10.0 \\
     2021/07/08 & 59403.8 & 10.0 \\
     2021/09/28 & 59485.6 & 3.8 \\
     2021/09/28 & 59485.6 & 4.4 \\
     2021/09/28 & 59485.9 & 0.4 \\
     2021/09/29 & 59486.0 & 0.1 \\
     2021/09/29 & 59486.0 & 0.7 \\
     2021/09/29 & 59486.0 & 3.8 \\
     2021/09/29 & 59486.2 & 0.0 \\
     2021/09/29 & 59486.2 & 7.5 \\
     2021/09/30 & 59487.8 & 4.2 \\
     2021/10/01 & 59488.1 & 10.0 \\
     2021/10/01 & 59488.7 & 6.5 \\
     2021/10/02 & 59489.0 & 10.0 \\
     2021/10/05 & 59492.1 & 10.0 \\
     2021/10/08 & 59495.7 & 0.4 \\
     2021/10/08 & 59495.7 & 6.7 \\
     2021/10/09 & 59496.4 & 4.7 \\
     2021/10/09 & 59496.7 & 7.8 \\
     2021/10/10 & 59497.1 & 1.7 \\
     2021/10/10 & 59497.2 & 10.0 \\
     2021/10/19 & 59506.8 & 4.3 \\
     2021/10/20 & 59507.0 & 10.0 \\
     2021/10/26 & 59513.5 & 10.0 \\
     2021/10/27 & 59514.6 & 9.2 \\
     2021/10/28 & 59515.2 & 7.6 \\
     2021/10/29 & 59516.2 & 10.0 \\
     2021/11/03 & 59521.6 & 9.3 \\
     2021/11/04 & 59522.3 & 5.8 \\
     2021/11/04 & 59522.6 & 9.1 \\
     2021/11/05 & 59523.3 & 7.9 \\
     2021/11/05 & 59523.6 & 8.4 \\
     2021/11/06 & 59524.0 & 10.0 \\
     2021/11/11 & 59529.2 & 10.0 \\
     2021/11/18 & 59536.5 & 10.0 \\
     2021/11/19 & 59537.6 & 10.0 \\
     2021/11/20 & 59538.1 & 10.0 \\
     2021/12/02 & 59550.5 & 10.0 \\
     2021/12/03 & 59551.9 & 1.8 \\
     2021/12/04 & 59552.0 & 3.7 \\
     2021/12/06 & 59554.8 & 4.6 \\
     2021/12/07 & 59555.2 & 10.0 \\
     2021/12/07 & 59555.8 & 3.0 \\
     2021/12/07 & 59555.9 & 1.3 \\
     2021/12/08 & 59556.3 & 10.0 \\
     2022/01/05 & 59584.7 & 6.5 \\
     2022/01/06 & 59585.2 & 1.9 \\
     2022/01/06 & 59585.3 & 10.0 \\
     2022/01/10 & 59589.8 & 4.5 \\
     2022/01/11 & 59590.2 & 10.0 \\
     2022/01/12 & 59591.6 & 8.7 \\
     2022/01/13 & 59592.0 & 10.0 \\
     2022/01/21 & 59599.9 & 1.1 \\
     2022/01/21 & 59600.3 & 7.4 \\
     2022/01/21 & 59600.6 & 8.6 \\
     2022/01/22 & 59601.7 & 1.9 \\
   \hline
    \end{tabular}
    \end{center}
    \addtocounter{table}{-1}
    \caption{-- {\it continued}}
\end{table}

\bsp    
\label{lastpage}
\end{document}